\documentclass[onecolumn,aps,preprint,showpacs,amsmath,amssymb]{revtex4}
\usepackage{graphicx}
\usepackage{latexsym}
\usepackage{xcolor}
\usepackage{dcolumn} 
\usepackage{bm}
\begin{document}
\newcommand{\eq}{\begin{equation}}                                                                         
\newcommand{\eqe}{\end{equation}}             
 
\title{Analytic solutions of the rotating and stratified hydrodynamical equations}

\author{Imre F. Barna$^{1}$ and L. M\'aty\'as$^{2}$}
\address{ $^1$ Wigner Research Centre for Physics, 
\\ Konkoly-Thege Mikl\'os \'ut 29 - 33, 1121 Budapest, Hungary \\ 
$^2$Department of Bioengineering, Faculty of Economics, Socio-Human Sciences and 
Engineering, Sapientia Hungarian University of Transylvania,  
 Libert\u{a}tii sq. 1, 530104 Miercurea Ciuc, Romania} 
\date{\today}

\begin{abstract} 
In this article we investigate the two-dimensional incompressible rotating and stratified, just rotating, just stratified Euler equations 
with each other and with the normal Euler equations with the self-similar Ansatz. 
There are analytic solutions available for all four models, 
for density, pressure and velocity fields, some of them are compound power-law dependent functions. 
In general the solutions have a rich mathematical structure. Some solutions show unphysical explosive 
properties others, however are physically acceptable and have finite numerical values 
with power law decays. For a better transparency we present some figures for the most complicated velocity and pressure fields.   
\end{abstract}

\maketitle

\section{Introduction}
There is no need to prove the evidence that both geophysics and meteorology 
have crucial importance for human society and civilization. Part of it is the special 
interest in science. Overwhelm problems in meteorology are hydrodynamical 
in origin. This statement is partially true for geophysics as well. 
  On the surface of Earth due to axial rotation and the additional gravity the question of stratified 
flows play an important role. Various hydrodynamical models of such kind for meteorology and geophysics 
can be found in numerous monographs like \cite{vallis,dol,zetl}.
	 
In this study we investigate the time-dependent dispersive self-similar solutions \cite{sedov,barenb} 
(not the blow-up type) of these kind of multidimensional Euler-type equations. The form of the 
original one-dimensional Ansatz reads as follows 
\eq
V(x,t) = t^{-\alpha} f(x/t^{\beta}) =  t^{-\alpha} f(\eta), 
\eqe
where $V(x,t)$ is the dynamical variable, $f(\eta)$ is the shape function with the reduced variable 
$\eta$ and $\alpha, \beta$ are the self-similar exponents.  Usually $\alpha, \beta >0$ 
present physically relevant power-law decaying physical solutions of the problem. 
This transformation is based on the assumption that a self-similar
solution exists, i.e., every physical parameter preserves its shape during the expansion. Self-similar solutions
 usually describe the asymptotic behavior of an unbounded or a far-field problem; the time t 
and the space coordinate x appear only in the combination of $\eta=x/t^{\beta}$. It means that 
the existence of self-similar variables implies the lack of characteristic lengths and times. 
The geometrical and physical interpretations of this Ansatz were exhaustively explained in 
all our former studies \cite{imre1,imre2,imre_book},  therefore we skip it here. 

This study is part of our long-time program which systematically goes over fundamental 
hydrodynamical systems. Till now we published about half a dozen papers  \cite{imre1,imre2} and a book chapter \cite{imre_book} in this field. 
To the best of our knowledge, there are no such time-dependent self-similar solutions known, presented 
and analyzed in the scientific literature for these systems.  
The structure of this paper is the following: to give a broader overview we investigate and compare 
the solutions of two dimensional rotating and stratified Euler equations with just stratified, just 
rotating and pure Euler equations. So four different flow systems will be discussed, 
We already published studies with the similar logic where several cases were investigated 
like the surface growth KPZ equation with numerous different noise terms \cite{imre_kpz} 
or the compressible one dimensional Euler equations where various equation-of-states were applied \cite{imre3}.    

\section{Theory}
To have a complex analysis for all four cases we present the corresponding original partial 
differential equation (PDE) systems, the applied Ansatz with the obtained self-similar exponents, the obtained coupled 
ordinary differential equation (ODE) system  and the solutions for the dynamical variables, 
the velocity and pressure fields (in two cases even for the densities). 
For a better transparency and for a clearer understanding we present figures for the most complicated 
solutions. These non-trivial shape functions and the corresponding final dynamical variables 
(velocity and pressure) are plotted and analyzed. We think that it is unnecessary to 
plot all shape functions and all dynamical variables for all four models for trivial solutions. 
\subsection{The rotating and stratified system} 
We start our study with the most complex flow. The stability or turbulence of such 
systems were extensively studied by 
Koba \cite{koba} and Davidson \cite{dav}.  According to the book of Dolzhansky \cite{dol} 
the rotating stratified fluid equations in two Cartesian dimensions in vectorial notation 
read as follows: 
\begin{eqnarray}
{\bf{\nabla}} {\bf{v}} = 0,  \nonumber  \\               
\rho_t + ({\bf{v}}{\bf{\nabla}} )\rho = 0, \nonumber \\
 {\bf{v}}_t  + ({\bf{v}}\nabla){\bf{v}} + 2{\bf{\Omega}}_0  \times {\bf{v}} =
 - \frac{\nabla p}{\rho_0} + \frac{G}{\rho_0}\rho,               
\label{eq} 
\end{eqnarray}
where ${\bf{v}}, \rho, p, {\bf{\Omega_0}}, G $ denote respectively the two-dimensional velocity field, 
density, pressure,  angular velocity and an external force (now gravitation) of the investigated fluid.  
In the following $\rho_0,$  is one physical parameter of the flow. 
For a better overview we use the coordinate notation ${\bf{v}}(x,y,t) = u(x,y,t),v(x,y,t) $ for the velocity 
and $p(x,y,t)$ for the scalar pressure field. To have a 
trivial rotation contribution we consider the 
${\bf{\Omega_0}} = (0,0,\Omega^z_0(x,y,t))$ angular velocity vector. 
The direct form, (coordinate form) of the equations are:  
\begin{eqnarray}
u_x + v_y  &=& 0, \nonumber \\
 \rho_t + u\rho_x + v\rho_y &=&0, \nonumber \\ 
u_t + uu_x + vu_y - 2v\Omega_0 &=& -\frac{p_x}{\rho_0},    \nonumber \\ 
 v_t + uv_x + vv_y + 2u\Omega_0 &=& -\frac{p_y}{\rho_0}  + \frac{G}{\rho_0}\rho,  
\label{eq2}
\end{eqnarray}
where the subscripts mean partial derivations with respect to time and spatial coordinates.  
(For the following three models we skip the vectorial form, and just write out all the coordinates.)  
 Let's consider the self-similar Ansatz for the variables in the form of:  
\begin{eqnarray}
\rho(x,y,t) = t^{-\alpha} f(\eta), \hspace*{3mm} 
u(x,y,t) = t^{-\delta} g(\eta), 
\nonumber \\
v(x,y,t) = t^{-\epsilon} h(\eta), \hspace*{3mm} 
p(x,y,t) = t^{-\gamma}i(\eta), 
\label{ans1}
\end{eqnarray}
with the new variable $\eta = \frac{x+y}{t^{\beta}}$. 
All the exponents $\alpha,\beta,\gamma,\delta, \epsilon,$ are real numbers. (Solutions with integer 
exponents are called self-similar solutions of the first kind, non-integer exponents generate self-similar solutions of the second kind.) 
The shape functions $f,g,h,i$ could be any continuous functions and will be evaluated later on.  
The logic, the physical and geometrical interpretation of the Ansatz were exhaustively analyzed in all our 
former publications \cite{imre1,imre2,imre_book,imre_kpz,imre3} therefore we neglect it.  

To have consistent coupled ODE system for the shape functions the exponents 
have to have the following values of 
\eq 
\alpha = 3/2, \hspace*{3mm} \beta = \delta = \epsilon = 1/2, \hspace*{3mm} \gamma = 1.
\eqe
Note, that all exponents have a fixed numerical value, which clearly defines the solutions. 
Each exponent is positive so the solutions are expected to be physical (will have power law decay at large times).   
It is important to emphasize, that only the ${\bf{\Omega}}^z_0 = \omega_0/t$ angular 
velocity function (which is trivial from dimensional consideration) leads the   
following clean-cut ordinary differential equation (ODE) system   
\begin{eqnarray}
f' + g' & = &0, \nonumber \\
-\frac{3}{2}f - \frac{1}{2}\eta f' + gf' + hf' &=&0, \nonumber \\ 
 -\frac{1}{2}g - \frac{1}{2}\eta g'  + gg' + hg' -2h\omega_0  &=& - \frac{i'}{\rho_0}, \nonumber \\ 
 -\frac{1}{2}h - \frac{1}{2}\eta h'  + gh' + hh' +2g\omega_0  &=& - \frac{i'}{\rho_0} + 
\frac{G}{\rho_0}f.    
\label{ode1}
\end{eqnarray}
From the first (continuity) equation we automatically get 
$ f + g = c_0$,
where $c_0 $ is proportional with the constant mass flow rate. Implicitly, larger $c_0$ means larger velocities. 
 From the first and second Eq. of  (\ref{ode1}) the ODE for the density shape function 
can be easily derived 
\eq
f'\left(c_0-\frac{\eta}{2}\right) - \frac{3f}{2} = 0.
\eqe
The solution is almost trivial 
\eq
f = \frac{c_1}{(2c_0-\eta)^3}
\label{f_eta}
\eqe
where  $c_1$ stands for the usual integration constant. 
The function is a shifted third order hyperbola with a singularity at  
$\eta = 2c_0$ for $\eta > 0 $ it is positive and strictly monotone decreases. 
The density has a power-law decay for large times which is physically desirable  
\eq
\rho(x,y,t) = \frac{1}{t^{\frac{3}{2}}} \cdot \frac{c_1}{ 2c_0 - \frac{(x+y)^3}{t^{\frac{3}{2}}  }} \simeq  \frac{c_1}{t^{\frac{3}{2}}}. 
\eqe

Extracting the fourth equation from the third one in \ref{ode1} the ODE 
for the shape function of the velocity component v can be easily given:  
\eq
h'(\eta - 2c_0) + h  - c_0\left(2\omega_0 + \frac{1}{2} \right) + \frac{Gf}{\rho_0} =0,  
\eqe
with the solution of 
\eq
h = \frac{\eta(-c_0-4c_0\omega_0)}{2c_0-\eta} - 
\frac{Gc_1}{\rho_0 (\eta-2c_0)^3} + \frac{c_2}{2c_0-\eta},   
\label{h_eta1}
\eqe
The function has a singularity at $\eta = 2c$ and it it strictly monotonous growing for all positive $\eta$ where 
$\eta > 2c_0$.  The solution is the sum of a shifted 
first and third order hyperbola.  All the parameters are 
responsible for the scaling and the shift of the singularity. 
It is straightforward to show that the asymptotic behavior of the velocity field is 
\eq
v(x,y,t)  \simeq  t^{-\frac{1}{2}},
\eqe
which makes it a physically acceptable solution. 

Adding the last two equations of Eq. (\ref{ode1}) the ODE of the pressure shape function can 
be obtained  
\eq
-\frac{2i(\eta)'}{\rho_0} + \frac{Gf(\eta)}{\rho_0}  + 4\omega_0 h(\eta) - c_0\left(\omega_0+\frac{1}{2}\right)= 0. 
\eqe
The solution can be easily evaluated with quadrature
 \begin{eqnarray}
 i =& 2\omega_0 ln(\eta-2c_0)(c_2 - 4\omega_0 c_0 -  \rho_0 c_0^2)+ 
\frac{Gc_1}{2(\eta - 2c_0)^2}\left(\frac{1}{2}-\omega_0 \right)  + \nonumber \\
  &\eta\left( \frac{1}{2}\omega_0 \rho_0c_0 - 4\rho_0\omega_0^2c_0 - \frac{c_0\rho_0}{4} \right)+ c_3. 
\label{i_eta1}
\end{eqnarray}
Figure (1) shows the pressure shape function for two different angular velocities giving qualitatively different 
curves. The integration constants $c_0, c_1, c_2, c_3$ play no  relevant role just 
shift and scale the results.  The key parameter is the angular velocity with the turning 
point of $\omega = 0.5$. In the case of $\omega_0 > 0.5$ there is 
a global maxima of the pressure.  Larger $\omega$ means quicker decay. Larger densities makes quicker pressure decays as well. 
 
To have a feeling about the general properties of the pressure, figure (2-3) present the ten-based logarithm of the 
solution for two different angular velocities. 
In both cases the pressure functions have clear asymptotic values.  
\begin{figure} 
\vspace*{-3cm} 
\scalebox{0.4}{
\vspace*{-3cm}
\rotatebox{0}{\includegraphics{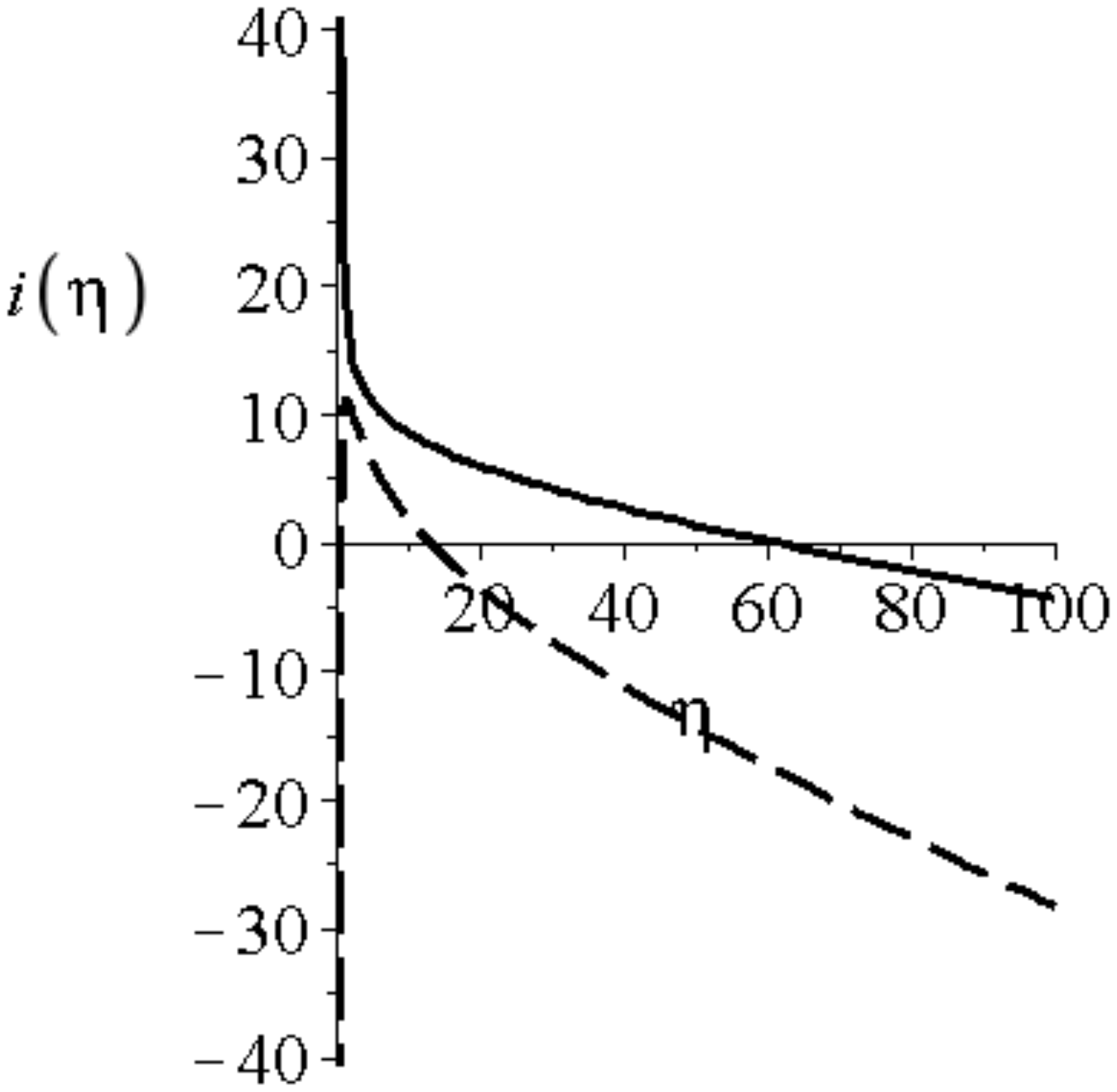}}}
\vspace*{-3cm}
\caption{Graphs of two different pressure shape functions Eq. (\ref{i_eta1}) where the common 
parameters are $G = 10,  \rho_0 = 1, c_0 = 1,  c_1 =3.25, c_2 = -3.1, c_3 = 15$. 
The solid and dashed curves are for $\omega_0 = 0.135 $ and $\omega_0 = 75 $, respectively.}
\label{egyes1}       
\vspace*{-4cm}
\scalebox{0.5}{
\rotatebox{0}{\includegraphics{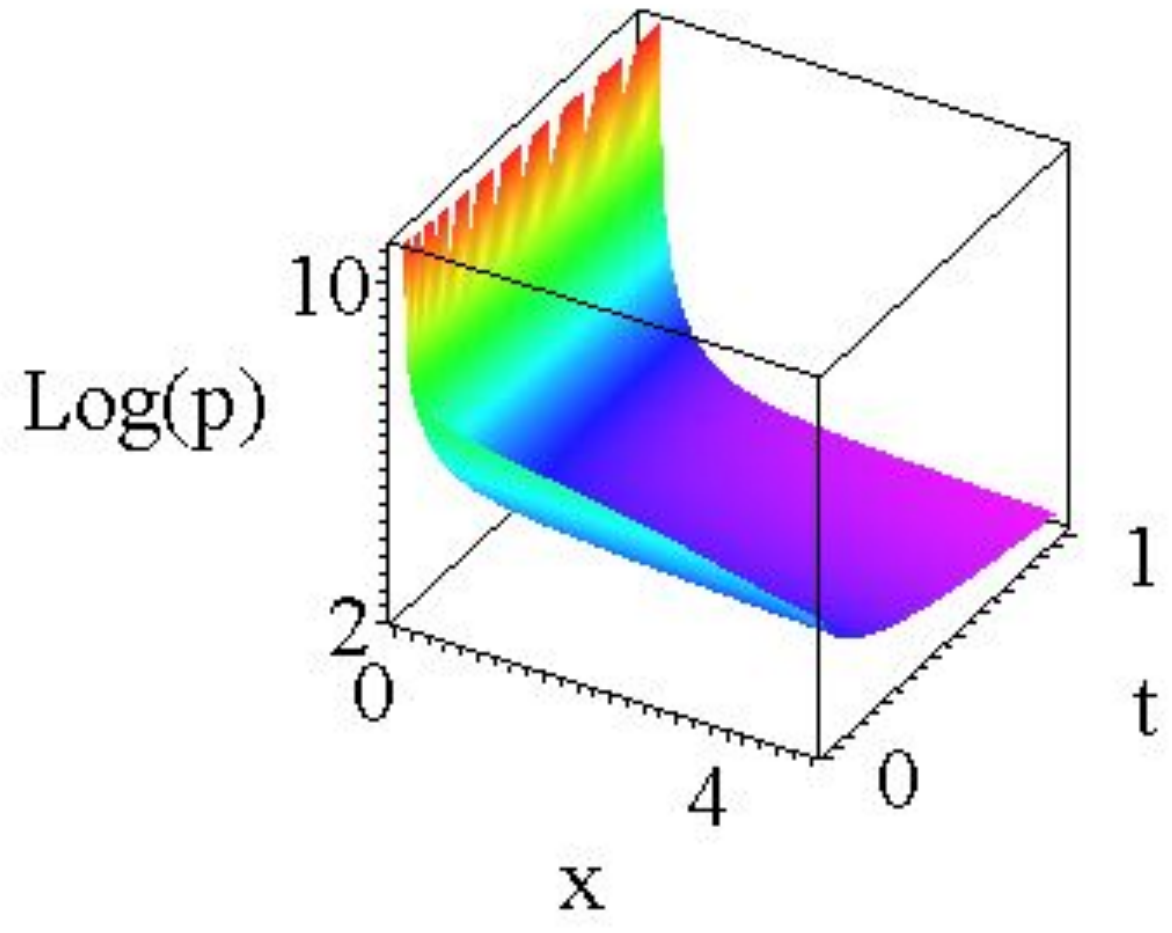}}}
\vspace*{-5.5cm}
\caption{The ten-based logarithm of the pressure $Log(p(x,y=0,t))$ for $\omega_0 = 0.135$ angular velocity,  
 all other parameters are given above.  }	
\label{kettes1}       
\vspace*{-3.7cm} 
\scalebox{0.53}{
\vspace*{-6cm}
\rotatebox{0}{\includegraphics{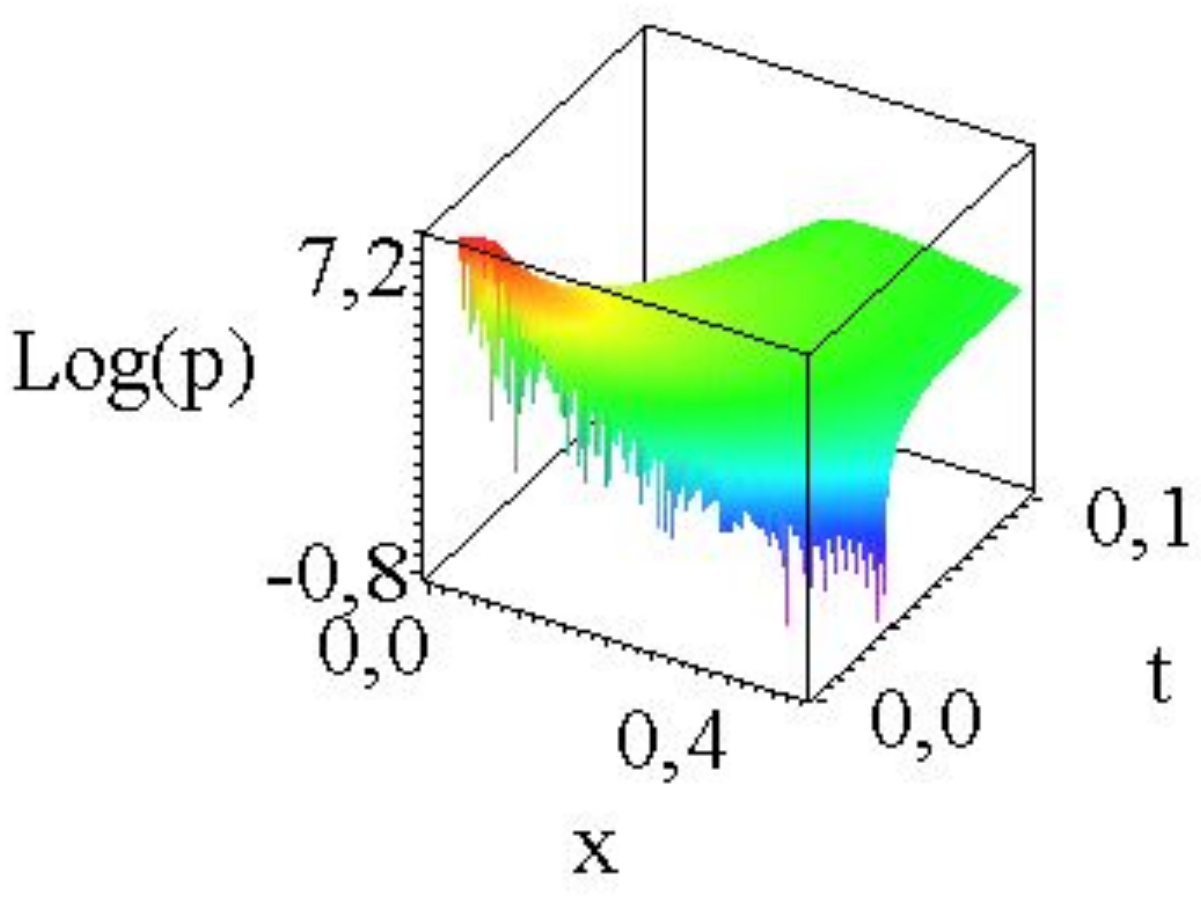}}}
\vspace*{-5.5cm}
\caption{ The ten-based logarithm of the pressure $Log(p(x,y=0,t))$ for $\omega_0 = 0.75$ angular velocity,   
 all other parameters are given above.  }	
\label{harmas1}       
\end{figure}
\subsection{The stratified system without rotation}
Now we consider the ${\bf{\Omega}} = 0 $ special case.   
The stratification and the change of density with the altitude has its importance 
in the Earth science. Numerous hydrodynamical questions in stratified flows (eg. turbulence) can be found in 
various textbooks \cite{ansorge,armenio,pedersen,vorop,chia,grim,hof}.    
Additional wave propagation issues were extensively studied, also see \cite{vil,ken}. 
 In 1975 Ono \cite{ono} presented algebraic solitary wave solutions for stratified fluids. 
An enhanced decrease of density of air with the altitude may lead to 
static stability which usually yields an increase of concentration of certain 
pollutants \cite{Pe2007,SzMa14,SzMaKeGh16}. 
The effect of temperature in the hydrodynamics of stratified flows may lead to specific convection 
phenomena even on small scales \cite{BaPoLoMa17}. 
There are discussed interesting aspects related to sedimentation in stratified flows as well 
by \cite{Da2016}. 

Our PDE system is now  
\begin{eqnarray}
u_x + v_y  &=& 0, \nonumber \\
 \rho_t + u\rho_x + v\rho_y &=&0, \nonumber \\ 
u_t + uu_x + vu_y  &=& -\frac{p_x}{\rho_0},    \nonumber \\ 
 v_t + uv_x + vv_y  &=& -\frac{p_y}{\rho_0}  + \frac{G}{\rho_0}\rho,  
\label{eq2}
\end{eqnarray}
where subscripts means partial derivatives with respect to time and both coordinates $x$ 
and $y$.  
The applied Ansatz is the following: 
\begin{eqnarray}
\rho(x,y,t) = t^{-\alpha} f(\eta), \hspace*{3mm} 
u(x,y,t) = t^{-\delta} g(\eta), 
\nonumber \\
v(x,y,t) = t^{-\epsilon} h(\eta), \hspace*{3mm} 
p(x,y,t) = t^{-\gamma}i(\eta), 
\label{ans2}
\end{eqnarray}
The corresponding ODE system is: 
\begin{eqnarray}
f' + g' & = &0, \nonumber \\
-(2-\beta)f - \beta\eta f' + gf' + hf' &=&0, \nonumber \\ 
 -(1-\beta)g - \beta \eta g'  + gg' + hg'  &=& - \frac{i'}{\rho_0}, \nonumber \\ 
 -(1-\beta)h - \beta\eta h'  + gh' + hh'   &=& - \frac{i'}{\rho_0} + \frac{G}{\rho_0}f.    
\label{ode2}
\end{eqnarray}
The slightly modified corresponding ODE system due to an undefined free self-similar exponent has a much 
larger degree of freedom.  So, in this sense all the exponents can be expressed with a fixed one (we may say with $\beta$)  
\eq 
\alpha = 2-\beta, \hspace*{3mm} \delta = \epsilon = 1-\beta , \hspace*{3mm} \gamma = 2(1-\beta).
\label{exp2}
\eqe
(We use $\beta$ as free parameter because it describes the common "spreading" property of 
all the dynamical variables, and now all "decay" parameters are free from 
each other.  So the "decays" of all variables can be studied independently.) 
The three ODEs for the shape functions can be determined with the logic mentioned above, 
\begin{eqnarray}
f'(c_0-\beta \eta) - f(2-\beta) & = &0,  \\
2h'(\beta \eta -c_0) - (1-\beta)c_0 + \frac{Gf}{\rho_0}&=&0,  \\ 
-\frac{2i'}{\rho_0}  + \frac{Gf}{\rho_0} + (1-\beta)c_0 &=&0.  
\label{odes}
\end{eqnarray}
All the solutions can be derived with quadrature 
\begin{eqnarray}
f &=& c_1(c_0-\beta\eta)^{\frac{\beta-2}{\beta}},  \label{f_eta2} \\
h &=&  -\frac{Gc_1(c_0-\beta \eta)^{\frac{\beta-2}{\beta}}}{2\rho_0(\beta-2)}
+ \frac{c_0 ln \left( [c_0-\beta \eta]^{\frac{\beta-2}{\beta}}\right)}{2(\beta-2)}
(1-\beta)+c_2,   \label{h_eta2}  \\   
i&=& -\frac{c_1G (c_0 - \beta\eta)^{\frac{2(\beta-1)}{2}}}{4(\beta-1)} + \frac{(1-\beta) c_0\rho_0}{2}\eta + c_3
 \label{i_eta2}.  
\end{eqnarray}
Note, that due to the free running self-similar exponent (now $\beta$) we got 
different kind of power-low dependent solutions, therefore this model has the richest mathematical structure.
To show the features of (\ref{f_eta2} - \ref{i_eta2}) we present and discuss some solutions with various $\beta$s. 
Figure (4) shows (\ref{f_eta2}) for six different exponents. 
Note, that we can get back all the usual power law functions, constant hyperbola and parabola as well. 
  Figure (5) presents the $h(\eta)$ shape functions. We present 5 different that kind of 
solutions, for reasonable $\beta$s.  
Figure (6) shows the shape functions for $i(\eta)$, there are 7 qualitative different functions exist as solutions. 
(We say that the typical exponent lies in the $[-4..4]$ range, for lot of physical systems 
this is restricted to the $[-2..2]$ interval.)  

Our decade long experience shows that  mainly the solutions with all positive exponents 
are physically relevant describing power-law dependent solutions. 
(Solutions with negative exponents usually have exploding properties at large 
times and space coordinates which violates mass, momenta or energy conservation.)    
Figure 6 - 8 present the tenth-based logarithm of the density, velocity and pressure for the 
common $\beta = 1/2$. Note, that all dynamical variables have a physically reasonable 
power-law decay for infinite times. 

\begin{figure} 
\scalebox{0.2}{
\rotatebox{0}{\includegraphics{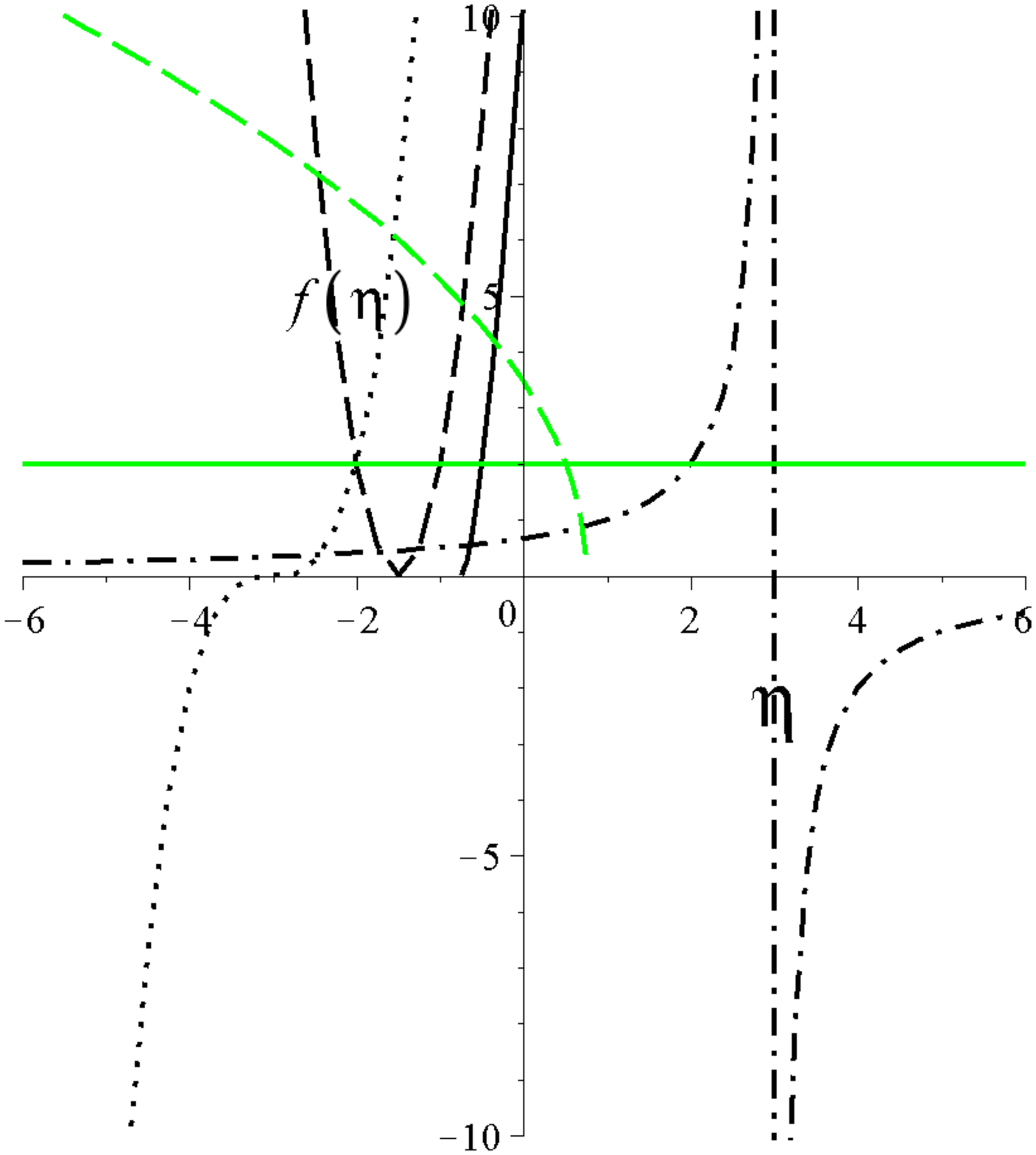}}}
\caption{The graphs of Eq.  (\ref{f_eta2}) the common parameters are $ c_0 = 4, c_1 =1.2$. 
The black solid, dashed, dotted and dash-dotted, the green solid and green dashed curves are for $\beta = -4, -2, -1, 1, 2, 4 $, respectively.}
\label{egyes}       
\scalebox{0.25}{
\rotatebox{0}{\includegraphics{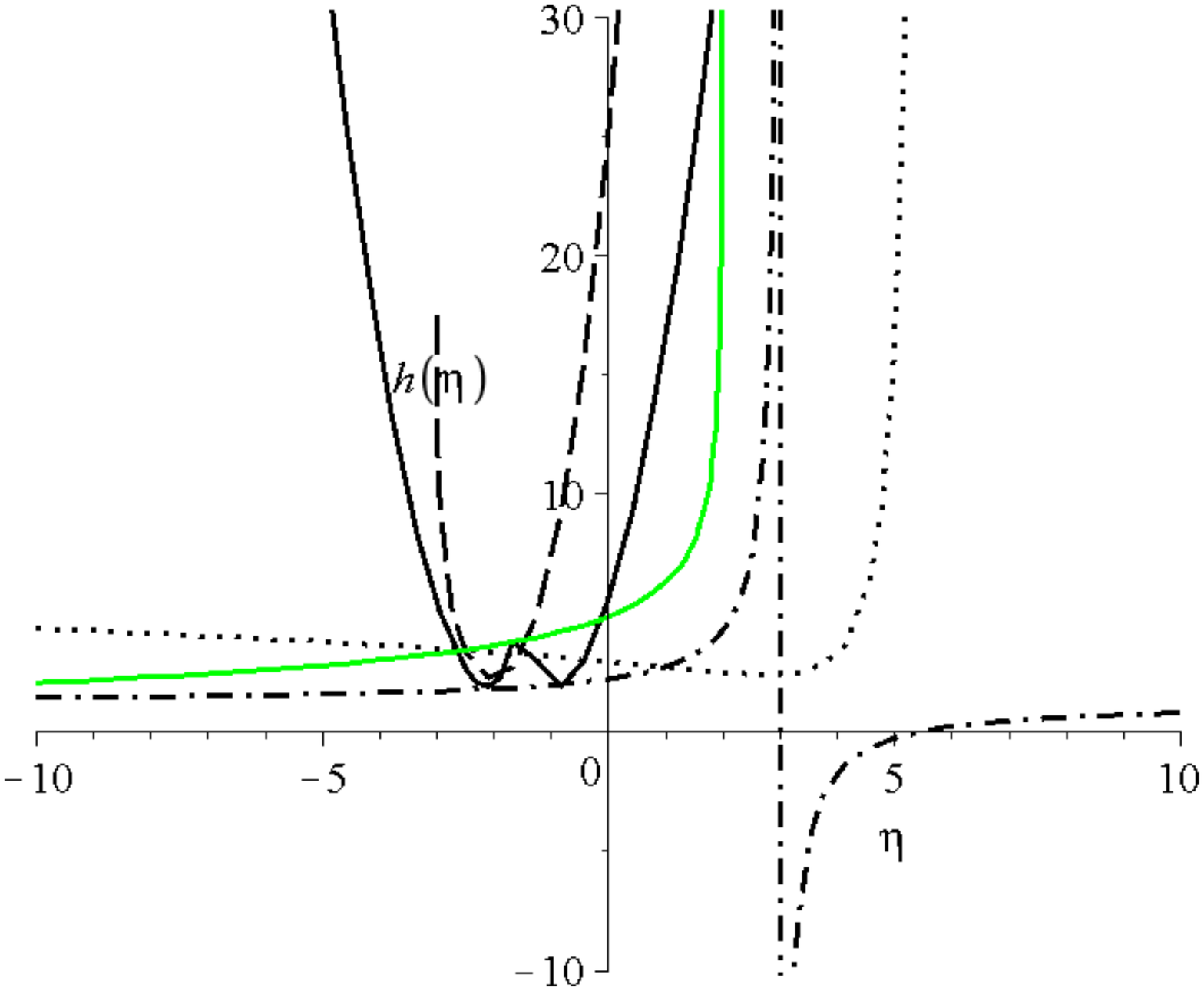}}}
\caption{The graphs of Eq.  (\ref{h_eta2}) the common parameters are $ G =10,\rho_0 = 1, c_0 =3, c_1 = 1.2, c_2 =1.2$. 
The black solid, dashed, dotted, dash-dotted and green solid curves are for $\beta = -2, -1, 1, 1.5, 2.5  $, respectively. }	
\label{kettes}       
\scalebox{0.22}{
\rotatebox{0}{\includegraphics{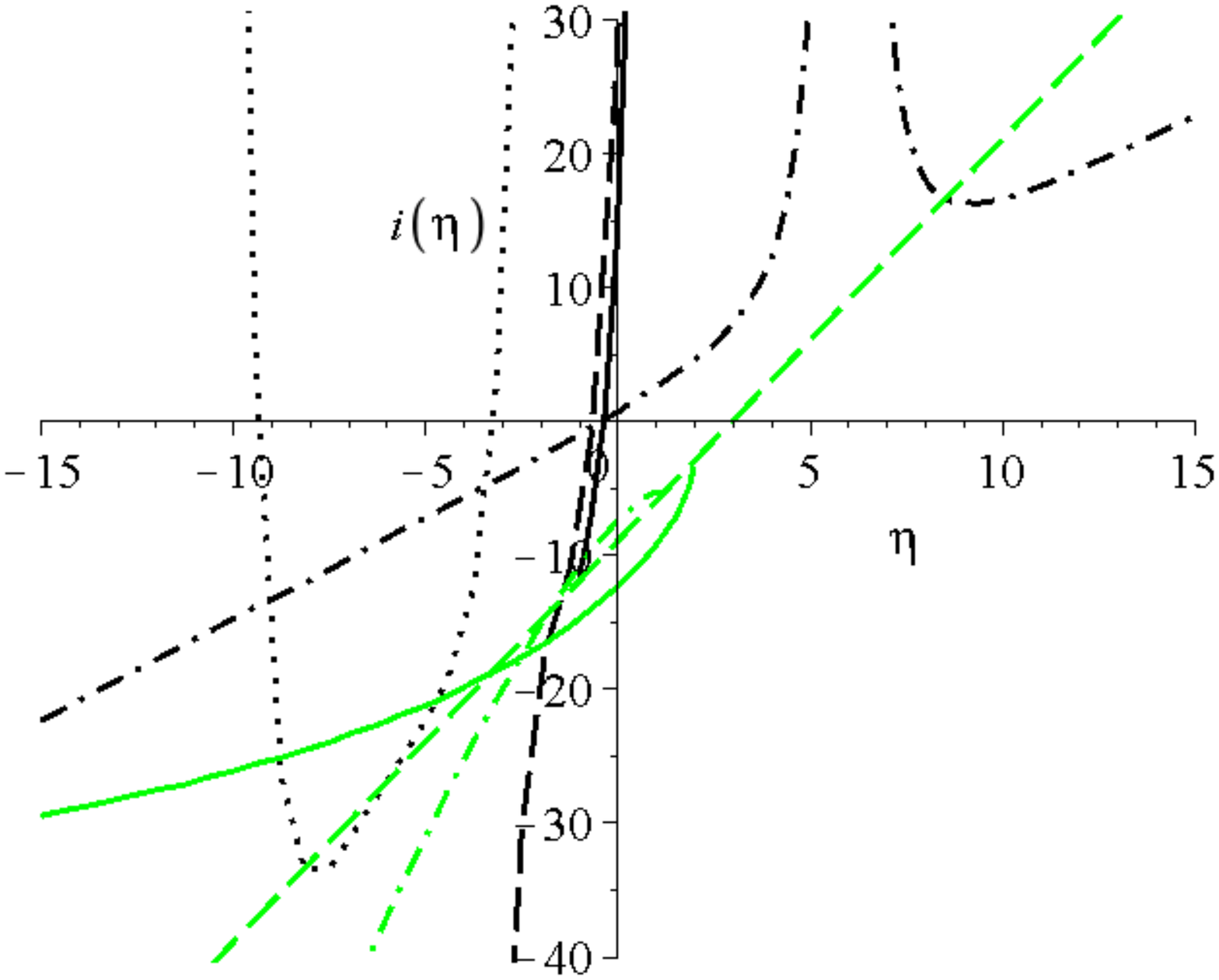}}}
\caption{The graphs of Eq.  (\ref{i_eta2})  the common parameters are the same as above with $c_3 = 0$.  
The black solid, dashed, dotted, dash-dotted and the green solid and green dashed  and green dashdotted curves are for 
$\beta = -3, -2, -0.5, 0.5, 1.5, 2, 2.5 $, respectively.   }	
\label{harmas}       
\end{figure}
 
\begin{figure}
 \vspace*{-2cm}
\scalebox{0.45}{
\rotatebox{0}{\includegraphics{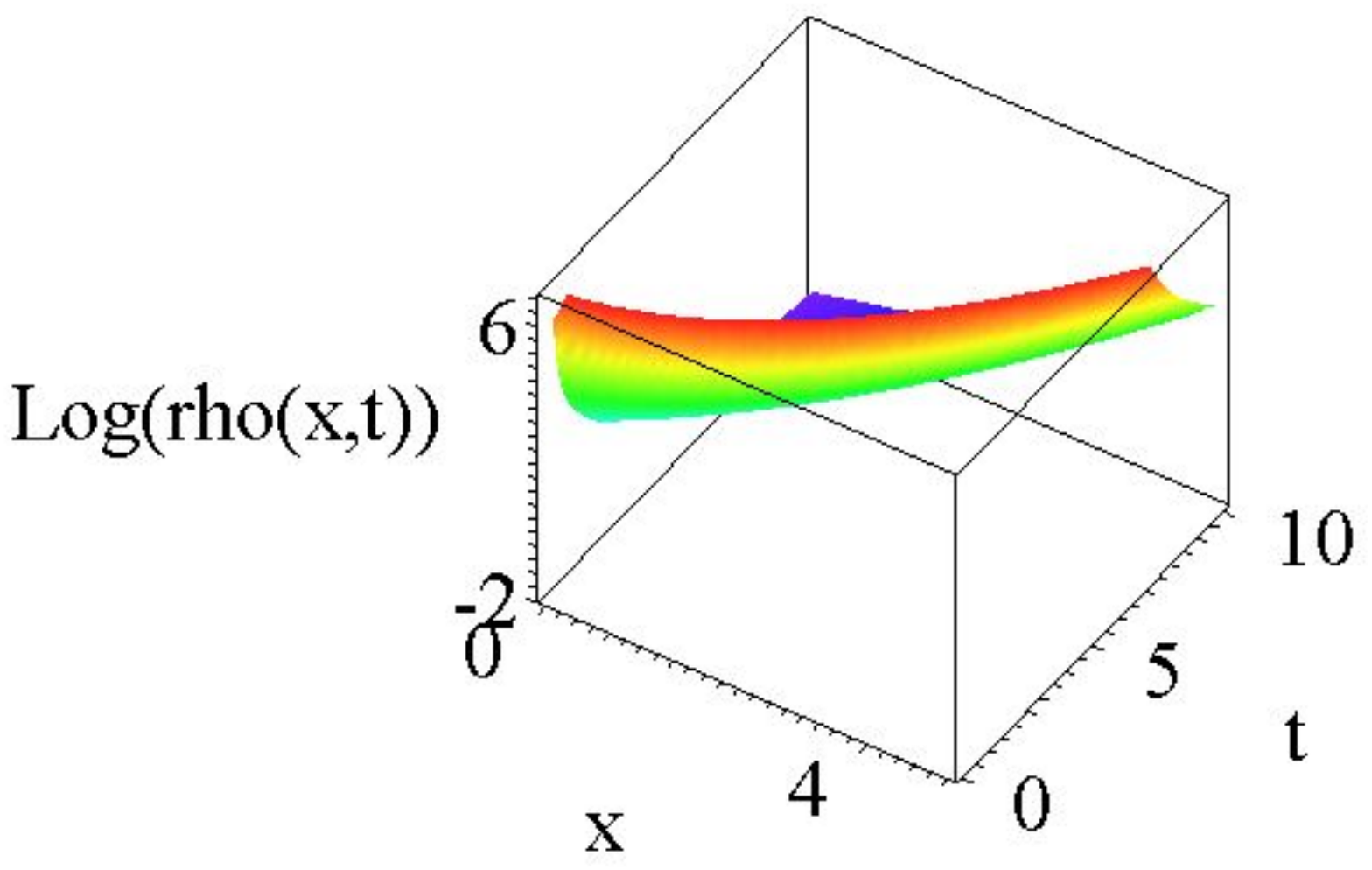}}}
\vspace*{-2cm}
\caption{The graph of the ten-based logarithm of the density function $\rho(x,y=0,t)$ for $\beta = 1/2$.}
\label{egyes}       
\scalebox{0.45}{
\rotatebox{0}{\includegraphics{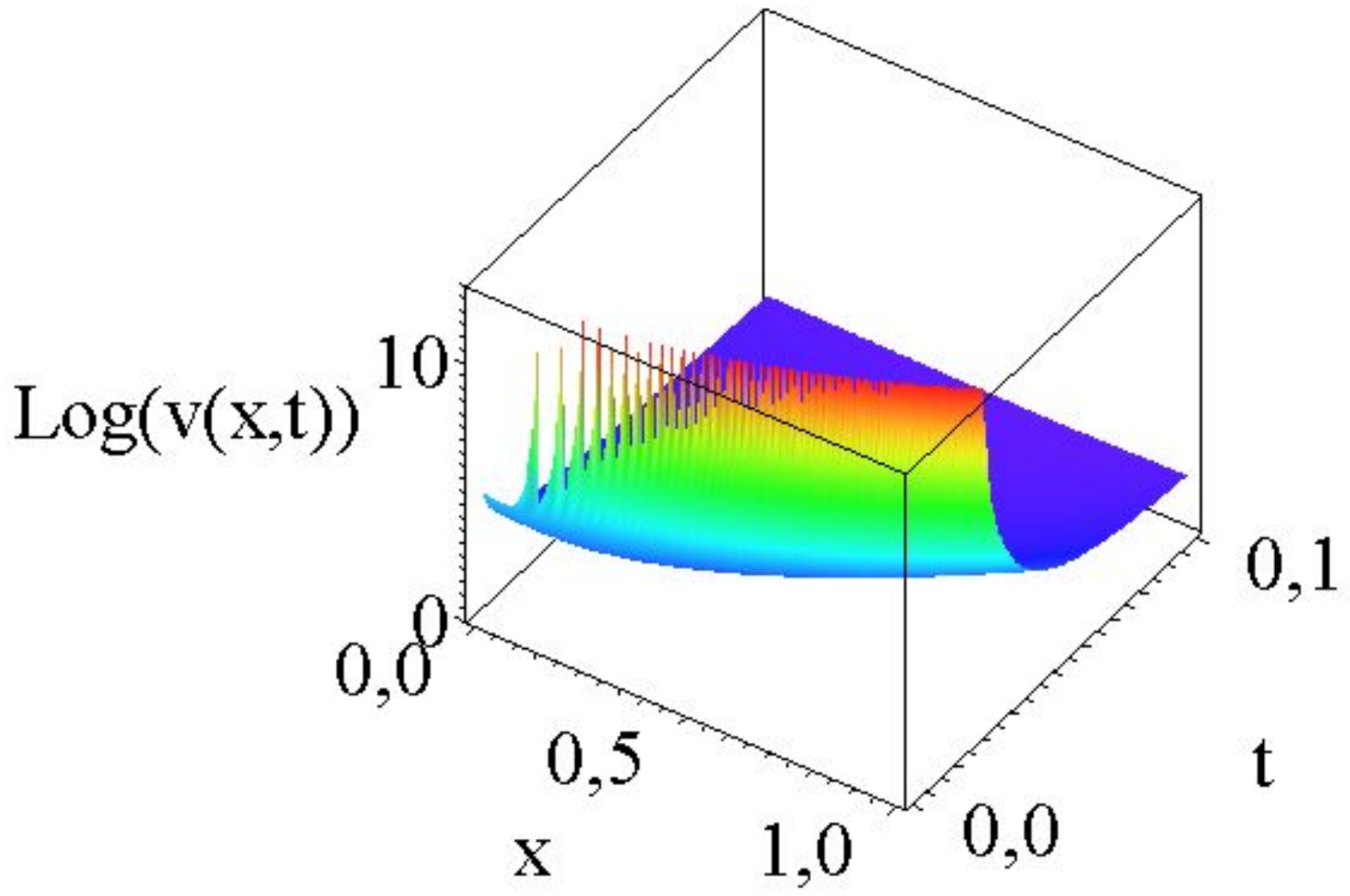}}}
\vspace*{-2cm}
\caption{The graph of the ten-based logarithm of the velocity function $v(x,y=0,t)$ for $\beta = 1/2$.}	
\label{v3d}       
\scalebox{0.45}{
\rotatebox{0}{\includegraphics{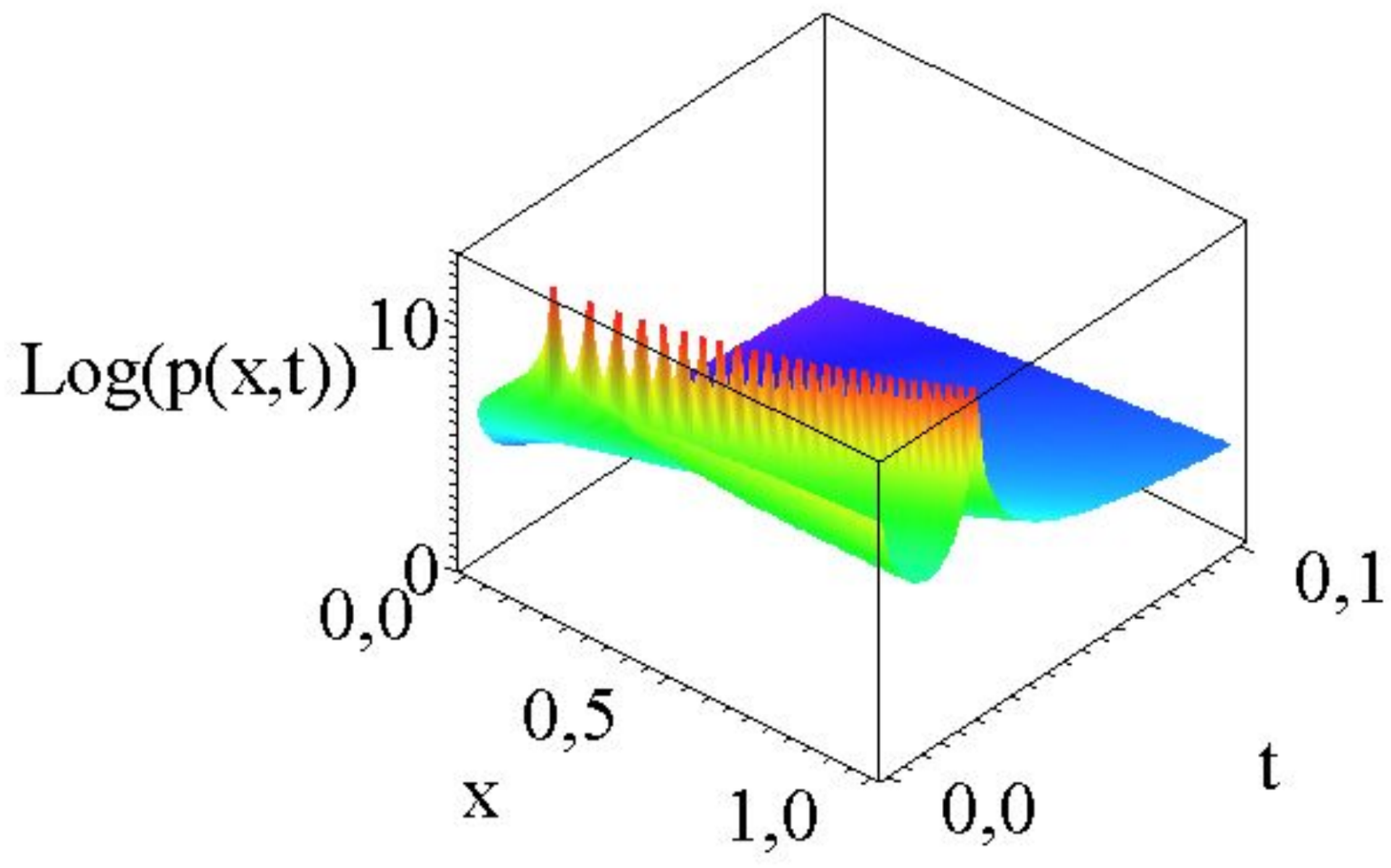}}}
\vspace*{-3cm}
\caption{The graph of the ten-based logarithm of the pressure function $p(x,y=0,t)$ for $\beta = 1/2$.}	
\label{p3d}       
\end{figure}

\subsection{The rotating system without stratification}
Rotating fluids are also relevant for science and engineering therefore the corresponding literature 
again enormous, without completeness we mention some of them  
\cite{boub,egb,eis,hopf2,vany,jean,keke,vadasz}.  
To investigate this case the complete second equation of (\ref{eq}) has to be neglected having the PDE system in the form of:
\begin{eqnarray}
u_x + v_y  &=& 0, \nonumber \\
u_t + uu_x + vu_y - 2v\omega_0 &=& -\frac{p_x}{\rho_0},    \nonumber \\ 
 v_t + uv_x + vv_y + 2u\omega_0 &=& -\frac{p_y}{\rho_0}  + G,  
\label{eq2}
\end{eqnarray}
So the number of the four unknowns is now reduced to three, namely to the velocity components $u,v$ 
and to the pressure $p$. To avoid contradiction among the exponents it is important to emphasize, 
that again only for ${\bf{\Omega}}^z_0 = \omega_0/t$ angular velocity function we get a clean-cut ODE system. 
The trial functions for the solutions now read 
\begin{eqnarray}
u(x,y,t) = t^{-\delta} g(\eta), \hspace*{3mm} 
v(x,y,t) = t^{-\epsilon} h(\eta), \hspace*{3mm} 
p(x,y,t) = t^{-\gamma}i(\eta). 
\label{ans2}
\end{eqnarray}
The relations among the exponents are the following: 
\eq
\beta = 2,  \hspace*{3mm}
\delta = \epsilon = -1 \hspace*{3mm} \gamma = -2. 
\eqe
The coupled ODE system is 
\begin{eqnarray}
g' + h' & = &0, \nonumber \\
 g - 2\eta g'  + gg' + hg' -2h\omega_0  &=& - \frac{i'}{\rho_0}, \nonumber \\ 
 h - 2\eta h'  + gh' + hh' +2g\omega_0  &=& - \frac{i'}{\rho_0} + G.    
\label{ode3}
\end{eqnarray}
The ODEs for one velocity field component and for the pressure field are: 
 \begin{eqnarray}
2h'(2\eta -c_0) - 2h - c_0(2\omega_0 -1)  + G &=& 0,  \\
-\frac{2i'}{\rho_0} + 4\omega_0h  G -c_0(2\omega_0+1) + G&=&0.
 \end{eqnarray}
There is no coupling between the variables. 
The corresponding solutions are 
\begin{eqnarray}
h &=&  c_1 \sqrt{c_0-2\eta} + \frac{G}{2} - \frac{c_0(2\omega_0-1)}{2}, \label{h_eta3}  \\
i  &=& \frac{1}{2} \rho_0\left( -\frac{4}{3}\omega_0c_1[c_0-2\eta]^{\frac{3}{2}}
+ \eta[ -2\omega_0 c_0 \{2\omega_0-1\} + 2\omega_0G -c_0\{2\omega_0 +1\} +G]
   \right) + c_2. \label{i_eta3}
\end{eqnarray}
The shape function of the velocity is a shifted square root function with negative argument. 
Note, the extra last positive shift term compared to  
the simple Euler case Eq. (\ref{h_eta4}) which is proportional to the angular velocity of the rotation $\omega_0$.   
The pressure shape function is a sum of a linear and an $\eta^{3/2}$ power law function with some shifts. 
Note, that the rotation is responsible to the first power law term,  
\begin{eqnarray}
v(x,y,t) &=& t^{-\epsilon} h(\eta) = t \left( c_1 \sqrt{c_0- \frac{2(x+y)}{t^2} }  \right)  + \frac{c_0(1-2\omega_0)+G}{2}, \\ 
p(x,y,t) &=& t^{-\gamma}i(\eta) = \nonumber \\ 
&=& \frac{t^2 \rho_0}{2} \left( -\frac{4}{3}\omega_0c_1\left[c_0-2 \frac{(x+y)}{t^2} \right]^{\frac{3}{2}}
+ \frac{(x+y)}{t^2}\left [    \tilde{C} 
  \right]\right)  + c_2,  
\end{eqnarray}
where 
\eq
\tilde{C} = -2\omega_0 c_0 \{2\omega_0-1\} + 2\omega_0G -c_0\{2\omega_0 +1\} +G.
\eqe
\subsection{No rotation and no stratification} 
This is the simplest system among the investigated four cases and this is the equation for 
the two dimensional incompressible ideal fluid as well. 
For completeness the starting PDE system is 
  \begin{eqnarray}
u_x + v_y  &=& 0, \nonumber \\
u_t + uu_x + vu_y   &=& -\frac{p_x}{\rho_0},    \nonumber \\ 
 v_t + uv_x + vv_y   &=& -\frac{p_y}{\rho_0}  + \frac{\rho}{\rho_0}G.  
\label{eq2}
\end{eqnarray}
The trial functions for the solutions are not changed from the previous case  
\begin{eqnarray}
u(x,y,t) = t^{-\delta} g(\eta), \hspace*{3mm} 
v(x,y,t) = t^{-\epsilon} h(\eta), \hspace*{3mm} 
p(x,y,t) = t^{-\gamma}i(\eta). 
\label{ans2}
\end{eqnarray}
The self-similar exponents remained the same too
\eq
\beta = 2,  \hspace*{3mm}
\delta = \epsilon = -1 \hspace*{3mm} \gamma = -2.  
\eqe
The ODE system is however a bit simpler 
\begin{eqnarray}
g' + h' & = &0, \nonumber \\
 g - 2\eta g'  + gg' + hg'    &=& - \frac{i'}{\rho_0}, \nonumber \\ 
 h - 2\eta h'  + gh' + hh'    &=& - \frac{i'}{\rho_0} + G.    
\label{ode3}
\end{eqnarray}
The decoupled ODEs for the velocity and for the pressure are also simpler,  
(note the missing terms with $\omega_0$)
 \begin{eqnarray}
2h'(2\eta -c_0) - 2h + c_0  + G &=& 0,  \\
-\frac{2i'}{\rho_0} -c_0 + G&=&0.
 \end{eqnarray}
The analytic solutions, after all, are almost trivial and read 
\begin{eqnarray}
h &=&  c_1 \sqrt{c_0-2\eta} + \frac{c_0+G}{2},  \label{h_eta4}   \\
i  &=& \frac{(G-c_0)\rho_0 \eta }{2}+c_2. \label{i_eta4}
 \end{eqnarray}
The velocity shape function is a square root function 
with shifted negative arguments which means that the 
function domain becomes negative. 
The shape function of the pressure is a simple linear function. Note the difference to Eq. (\ref{i_eta3}) is 
due to the rotation $\omega_0$. 
For completeness the final field variables are
\begin{eqnarray}
v(x,y,t) = t^{-\epsilon} h(\eta) = t \left( c_1 \sqrt{c_0- \frac{2(x+y)}{t^2} }  \right)  + \frac{c_0+G}{2}, \\ 
p(x,y,t) = t^{-\gamma}i(\eta) = t^2  \left(  \frac{(G-c_0)\rho_0}{2} \frac{(x+y)}{t^2} \right)+ c_2.
\end{eqnarray}
Notice, that both dynamical variable have no decay property to large times, therefore we consider them 
unphysical and skip to present additional figures. 
From physical considerations we may calculate the total kinetic energy term which is proportional to the volume integral of 
$ \int_V \frac{\rho_0}{2} [u(x,y,t)^2 +  v(x,y,t)^2]dx dy $ this quantity should however has a time decay at infinite times 
(where V means the volume of the dynamics). 
 
\section{Summary} 
We investigated the two-dimensional incompressible rotating and stratified, just rotating, just stratified Euler equations 
with each other and with the normal Euler equations applying the self-similar Ansatz. To emphasize 
the scientific relevance of these equations and disciplines we mentioned numerous textbooks and 
monographs which were written in the recent decades. We found analytic solutions for all dynamical 
variables of all four models. The solutions of the rotating stratified and the stratified flows are much 
more complex than the last two one, therefore we presented additional figures to enlighten the details. 
Overall the physically relevant, power-law time decaying solutions were emphasized. We think that due to the lack 
of higher order viscous terms in the Euler equations all solutions are quite simple contains 
no additional internal finer structure e.g. some waves or oscillations.  
We would like to publish this manuscript just as a precursor of planned later studies with more 
complex materials of  viscous fluids.
   
 
\end{document}